# Transition from multiphoton to tunneling ionization in the process of high harmonic generation in solids

Shuai Wang[1,2], Xinkui He[1,2,3,a], Yueying Liang[1,2], Yabei Su[1,2], Shaobo Fang[1,2], and Zhiyi Wei[1,3,b]

[1] Beijing National Laboratory for Condensed Matter Physics, Institute of Physics, Chinese Academy of Sciences, Beijing 100190, China

[2] University of Chinese Academy of Science, Beijing 10049, China

[3] Songshan Lake Materials Laboratory, Dongguan 523808, Guangdong, China

High harmonic generation in solids is becoming an important method for strong field solid state physics research. The power scale relationship between high harmonics and the driving laser is investigated both experimentally and theoretically. Results of the power scale dependence clearly divided the interaction into two regimes. The modification of the bandgap by intense laser proved to be very important for theoretically reproducing the experimental result. Combining with the Keldysh theory analysis, the harmonic generation process is found to be transitioned from multiphoton excitations to the diabatic tunneling.

High-order harmonic generation (HHG) in gas has provided an important source for coherent extreme ultraviolet radiation and ultrafast attosecond pulse generation. Comparatively HHG in solids produces radiation with lower photon energy and longer pulse duration[1-6]. However, inherent connection with the strong field interaction makes it a new method for studying strong-field condensed-matter physics. High harmonics from solids are now used to investigate the essential properties of crystals in the strong laser field, such as the band dispersion[7-9], the berry phase[10,11], and the periodic arrangement of atoms in crystals[12].

As an extreme nonlinear process, the yield of the emitted harmonic radiation shows highly nonlinear dependence on the driving laser fields [1,3,11,13]. Not like the case in the perturbative limit, the $n^{th}$ harmonic yield to the scale as the $I^n$, almost a fixed scale number can be obtained for the power-law of the harmonic yields as a function of driving field intensity[1,13], which shows the non-perturbative character of the HHG process. High harmonic generation in solids is often theoretically described with two different process called inter-band radiation which is from the recombination of the electron and hole between the conduction and valence bands, and intra-band radiation from laser-driven oscillation of electron and hole within bands [14-19]. The first step of whole process is the strong field ionization of the electron and forming electron hole pairs in the bands, which defines the non-perturbative nature of the process.

The strong field ionization process is very well described by the famous Keldysh model[20]. A dimensionless parameter $\gamma$ defined as the ratio between tunneling time with the period of the driving light field, was introduced to classify the regimes of the ionization. In the multiphoton regime, the $\gamma \gg 1$, the electrons are excited from valence band to conduction band by absorption of multiphoton energy from laser field. In the diabatic tunneling regime, the $\gamma \sim 1$, the potential becomes narrower and the electrons getting the multiphoton energy easily escape from the potential well by tunneling. In the regime of adiabatic tunneling $\gamma \ll 1$, an electron tunnels through the potential barrier without changing its energy. The signature feature of the tunneling ionization is the laser field modification of the potential and lower the barrier. However, the modification of the periodic potential of solids is normally not included in theoretic analysis of HHG in solid. As the increase of the driving laser field, the shape of the potential and the bandgap between the valence and conduction band are modified, which modifies the process of the HHG. On the other hand, the generated harmonics carry the information of the potential. Considering the field-induced band structure modification, picometre-scale imaging of valence electron has been extracted from generated high harmonics[21]. As an emerging method for strong field solid state physics investigation, it is important to know which interaction regime this process belongs to, and also critical to understand what effect the potential modification plays. The power scale relationship between the harmonics and the driving laser has played an important role for understanding the generation process in perturbative regime. Whether it can also be used to understand the no perturbation process, it is necessary to further study the power scale dependent of the HHG in solid.

In this Letter, the power scale dependence between the high-order harmonic and the driving laser has been investigated both experimentally and theoretically. Polycrystalline $SiO_2$ film was driven by sub-10 femtosecond in the experiment. Two different slopes in the power scale dependence curve clearly divided the interaction into two regimes. Simple theoretic model by using the laser field modified bandgap in the semiconductor Bloch equation (SBE) shows the result attributed to the bandgap modification. The Keldysh theory analysis revealed the interaction process was transitioned from multiphoton excitations to the diabatic tunneling.

A Ti:sapphire laser with the repetition rate of 1 kHz, pulse duration of 35 fs and the maximum pulse energy of 7 mJ was used in the experiment. The laser pulse was further compressed using a neon-gas-filled hollow-core fiber spectrum expander and chirped mirrors dispersion compensator. Finally, laser pulse with duration of 9.49 fs and energy of 0.95 mJ was used for HHG. A spherical mirror with the focal length of 1.3 m was used to focus the laser to a free-standing polycrystalline $SiO_2$ film with the thickness of 100nm, which was mounted on a rotary table. A tunable aperture was put before the focus mirror to tune the intensity at the laser focus. The generated harmonic spectrum was measured by a homemade flat-field

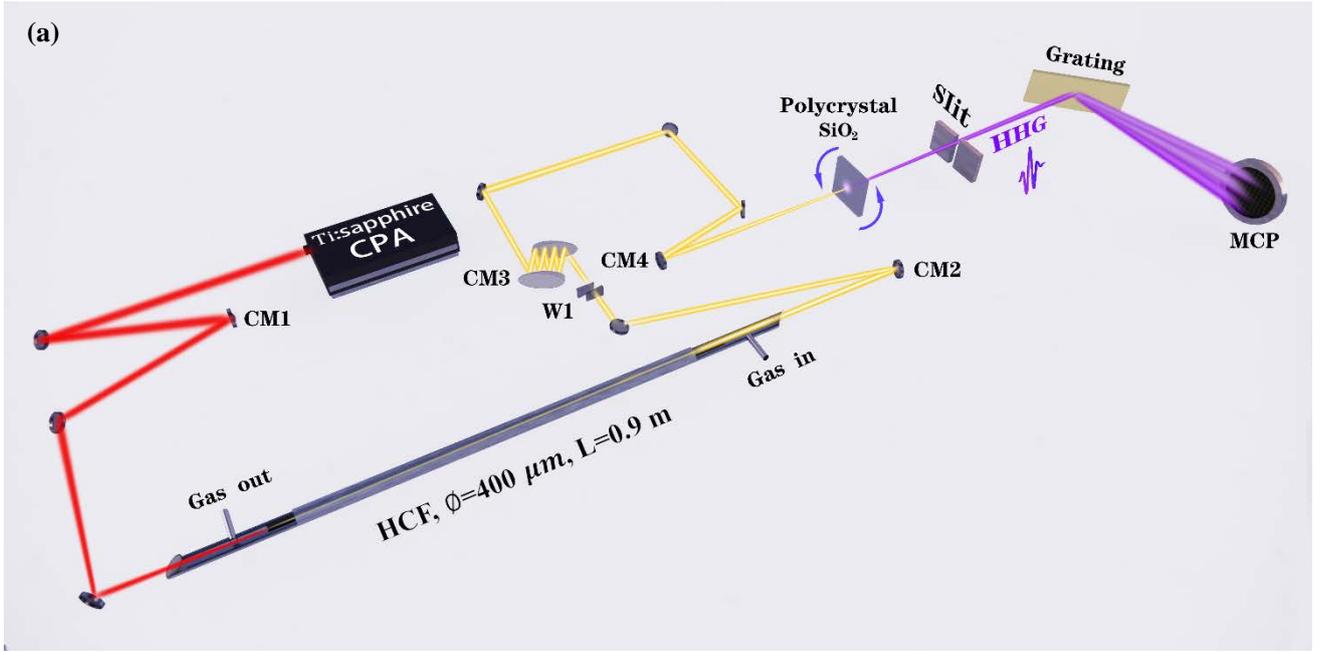

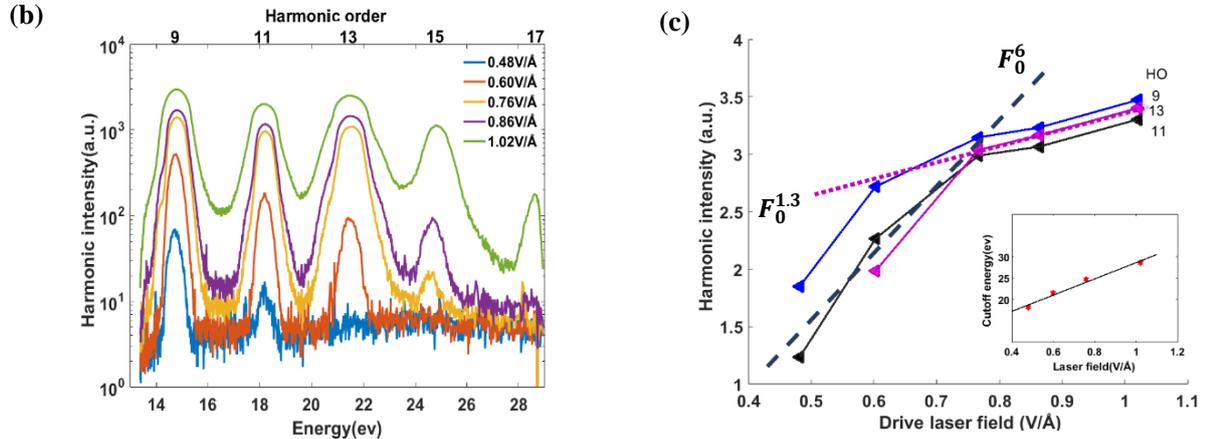

Fig 1 (a) Schematic of experimental set up. CM1-2, curved mirror with focal length f1=2.5m and f2=2m; W1, wedges; CM3, chirped mirrors; CM4, curved mirror with focal length f4=1.3m. (b)Measured high harmonics in polycrystalline $SiO_2$(100nm) as a function of laser field strengths. (c)Yield dependence for the 9$^{th}$, 11$^{th}$ and 13$^{th}$ harmonics. The dash and dot lines are the fits of the experimental data to a power law $I^p$. The harmonic intensity grows rapidly in dash lines and slowly in dot lines with field amplitude. The inset figure shows the cutoff photon energy $E_{cutoff}$ scales with the field amplitude $F_0$(red hexagram). The black line indicates the linear fitting.

spectrometer consisting of a grating and microchannel plate. The harmonic spectra and spatial distribution on phosphor screen of MCP is recoded by a charge-coupled device (CCD) camera. The diagram of the experiment set up is shown in Fig.1.a.

The high-order harmonic spectrum from the bulk of the $SiO_2$ for different driving field intensity is shown in Fig.1b. The isotropic structure of polycrystalline $SiO_2$ determines only the odd harmonics was observed. The harmonic intensity and the cutoff energy increase with the driving field strength. The harmonic yield as a function of driving laser field intensity is shown in Fig. 1c. It clearly shows that there are two different regions with different power scale. At lower field strength, the harmonic intensity rapidly increases with the intensity of driving laser field, whereas at higher intensities, the growth scaling suddenly slows down. Power-law fitted to the data is obtained, the 9$^{th}$,11$^{th}$,13$^{th}$ harmonics firstly follow rapid scale around $F_0^6$, and then follow almost the same slow scale of $F_0^{1.3}$. The transition between the two power-scales is around $F_0 = 0.75$ V/Å. Inset in Fig.1c shows the cutoff photon energy undergoes an increasing linear scaling. Following this the largest characteristic distance $n_{max}a$ that electron wave packet can reach is about $4a$, where $a = 5$ Å is the lattice constant.

To understand the physics behind our experiment, the HHG process is simulated using SBE. The SBE equation with two-band system can be written as,

$$\frac{\partial \pi(K,t)}{\partial t} = -\frac{\pi(K,t)}{T_2} - i\xi(K,t)[n_v - n_c]e^{-iS(K,t)} \quad (1)$$

$$\frac{\partial n_v(K,t)}{\partial t} = -i\xi^*(K,t)\pi(K,t)e^{iS(K,t)} + c.c. \quad (2)$$

$$\frac{\partial n_v(K,t) + n_c(K,t)}{\partial t} = 0 \quad (3)$$

where $n_m$ (m=c, v) is the population of the conduction band and valence band, $\epsilon_g = E_c - E_v$ is the bandgap, $S(K,t) = \int_{t_0}^{t'} (\epsilon_g(K + A(\tau))) \, d\tau$ is accumulated phase of electron-hole during the motion in the corresponding conduction and valence bands. However, the actual energy bands of the electron-hole pair have modified by the strong laser field. Therefore, it is necessary to take into account the energy bands varying with the laser field in the calculation. $\xi^*(K,t) = F(t)d(K + A(t))$ is the Rabi frequency, and $d(k) = i\int d^3x u_{v,k}^*(r)\nabla_k u_{c,k}(r)$ is the transition dipole moment, with $u_{m,k}(r) = \sum_G u_{k,G}^m \exp(i(G)r)$ the periodic part of the Bloch function. $K = k - A(t)$ is the shifted crystal momentum with the vector potential $dA(t)/dt = -F(t)$, and the first Brillouin zone is also shifted to $\overline{BZ} = BZ - A(t)$. $T_2 = 0.5T_0$ is the dephasing-time describing the coherence time between the valence and conduction bands. The total emission of harmonics can be calculated by the Fourier transform of the intra- and interband currents. As the intraband radiation dominates the harmonic generation process in SiO2[3, 10, 21-23], the inter-band current is neglected in our simulation.

$$j_{ra}(t) = \sum_{m=c,v} \int \mathbf{v}_m(k) n_m(k,t) d^3k \quad (4)$$

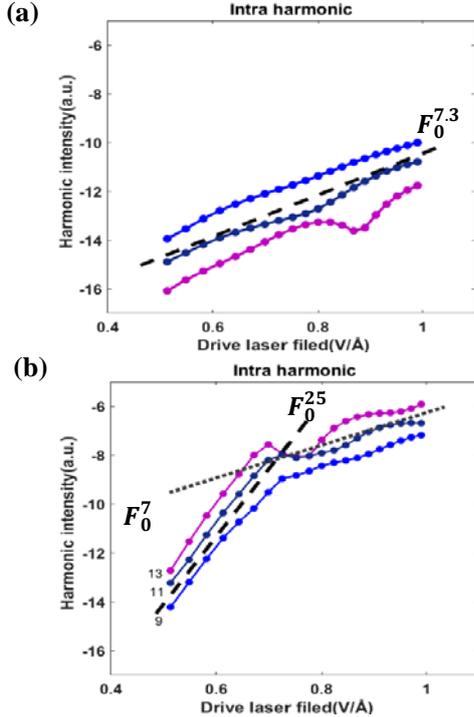

Fig 2 Simulated intraband harmonics intensity versus $F_0$ without(a) and with(b) band structure modification. The wavelength of the driving laser field used in the theoretical simulation is consistent with the experiment centered at 780 nm. The blue, black and purple solid lines show the 9, 11th and 13th harmonic orders intensity versus laser field strength. The dashed line and the dot line are the fitted power law from the simulated data.

The band structures $E_m(m = c, v)$ are calculated by the effective potential in a laser field ($\hbar w_L = 1.58$ eV) of different $F_0$. Within the picture of laser field modified band structure, the radiation of HHG can be simulated using SBE in different laser field strength $F_0$. We expand the energy $E_m(k) = \sum_{i=0}^{i_{max}} \varepsilon_{m,i} \cos(iak)$ in the $m$ th band on crystal momentum $k$, where $i_{max}a$ is the largest characteristic distance with lattice constant $a = 5$ Å [1, 3]. The band structure of crystal $SiO_2$ was used to substitute that of polycrystalline $SiO_2$ with the spatial harmonics only to the largest characteristic distance $i_{max}a = 4a$.

Figure 2 shows the calculated harmonic intensities as a function of driving laser field strength on a logarithmic scale. The power scale of harmonics calculated without the field-induced band structure modification are almost linear with a same logarithmic scale in the whole field strength range (Fig.2a). For the field modified band structure case, two different power scale for different field intensity can be found. Fit of the simulation curve to a power law of $F_0^p$. At lower field intensities, all the calculated harmonics are highly nonlinear with a similar intensity dependence $\propto F_{peak}^{25}$, whereas an obviously smaller exponential dependence of 7 is obtained for higher field intensities. The calculated power scale is distinctly bigger than that from experiment data, which is understandable since the single-electron approximation and one-dimensional potential were used in the simulation and the effect of interaction between electrons was not included. However, for both the experiment and simulation, the laser field range for power scale dependence transition from lower two higher is striking agree, which is around $F_0 = 0.75$ V/Å. This implies the existence of different mechanism for two field intensity ranges which is the result of the field modification of the band structure. The transition between two mechanisms was happened with field intensity around 0.75 V/Å.

To reveal the change of the electron dynamics with the increase of the diving field intensity, we carefully investigated how the periodic potential and band structure $E_m$ is influenced by the high intense laser field. The potential is expanded in the plane-wave basis $exp(iGr)$ with $G = N * 2\pi/a, N \in Z$. Assume laser field $F(t) = F_0 \sin(w_L t)$ with the vector potential $-dA/dt = F$, the total potential is formulated [21, 24] in KH frame.

$$V\left(r + \int A dt\right) = \sum_{G,n} V_G J_n \left(\frac{GF_0}{w_L^2}\right) \exp(iGr - inw_L t) \quad (5)$$

Here $J_n$ is the Bessel function of the first kind of order $n$. For $n = 0$, the time independent terms of the total potential denote an effective crystal potential $V_{eff}(r, F_0) = \sum_G V_G J_0 \left(\frac{GF_0}{w_L^2}\right) \exp(iGr)$, which are modified by the laser field and significantly influences the band structure in solid. For $n \neq 0$, the time dependent terms $\sum_G V_G J_{n\neq 0}\left(\frac{GF_0}{w_L^2}\right) \exp(iGr - inw_L t)$ in equation describe the transition among the states of the effective potential effective potential[25], which can be represented by SBE in detail.

If laser field $F_0 = 0$, the field-induced effective

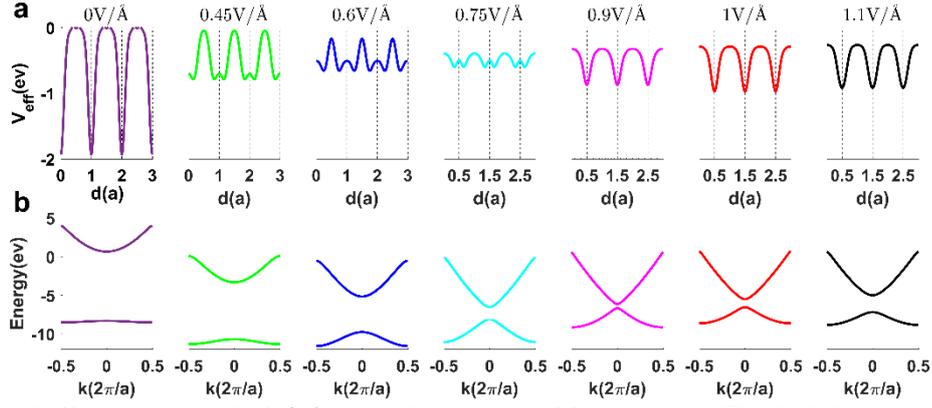

Figure 3 (a) Calculated effective potential of $SiO_2$ with the increase of field strength. The dash lines represent the valley bottom of the effective potential. (b) Calculated band structure for the last valence and the first conduction bands for the corresponding field strengths.

potential $V_{eff}(r, F_0) = \sum_G V_G J_0(\frac{GF_0}{w_L^2}) exp(iGr)$ degenerates into laser field-free crystal potential $V(r) = \sum_G V_G exp(iGr)$. To describe the field-free potential accurately, we should consider as many orders of reciprocal-space vector as possible. In this paper, one dimensional potential of crystal $SiO_2$ along Γ–M direction was expanded in plane wave $V(x) = \sum_G V_G exp(iGx)$, which has been used to analyze interband tunnelling and the quantum-beat of the polarization response [26]. When the reciprocal-space vector is expanded to order $of$ N = 8, the fitting of R-square is about 95%, which is enough to describe the field-free potential.

Substituting $V_G$ into $V_{eff}(x, F_0)$, one dimensional field-induced effective potential $V_{eff}(x, F_0) = \sum_G V_G J_0(\frac{GF_0}{w_L^2}) exp(iGx)$ is obtained. The electrons can be driven as free particles by an external laser field in the zero effective potential. However, it is difficult to meet this condition since it is almost impossible to have $J_0(\frac{GF_0}{w_L^2}) \approx 0$ in the specific laser field $F_0$ for different order reciprocal-space vectors $G$, which implies the electrons cannot be taken as complete free particles in an external laser field.

The band structure is calculated using the effective potential in a laser field ($\hbar w_L = 1.58$ eV) for different intensity $F_0$ is shown in Fig. 3a. Two distinct intensity regions can be seen. With the increase of the laser field, the modulation of the effective potential can be found in both amplitude and position of peaks and valleys. The absolute value gradually decreasing to minimum at critical field strength of $F_0 = 0.75$ V/Å. The minimums of effective potential are at $d = N * a, N \in Z$. When the field strength $F_0 = 0.75$~1.1 V/Å, the valleys' intensities gradually increasing in reverse makes the valleys finally becoming peaks. The new valleys of effective potential are at $d = (N - 0.5) * a, N \in Z$. The absolute values of the new valley bottom increase quickly when the laser field $F_0 = 0.75$~0.9 V/Å. However, the effective potential changes slowly as the laser field strength gets higher.

The modification of band structure with the increase of the laser field strength is shown in fig. 3(a). As the laser field increases, the modified bandgap decreases quickly and reaches minimum at the field strength $F_0 = 0.9$ V/Å. Since the bandgap is not close, the band dispersion does not approach that of the free electron. As the laser field strength further increases, the bandgap gradually opens. This implies that the strong laser field changes not only the trough value of effective potential but also the position of effective potential, both can affect the band structure of solid.

The modulation of the effective potential and modification of the bandgap consequently affected the generation of the harmonics. To understand how this effect works, the calculated population of the conduction band is shown in fig. 4(red line). The growth rate of the electron population in the conduction band clearly shows three different value which divided the laser field in three regimes. For the laser field below 0.75 V/Å ,the population increases very fast with the increase of the driving laser field intensity. For the laser field strength from 0.75 to 0.9 V/Å, the growth rate of the conduction population is becoming smaller, and almost constant in the small range. For the laser field stronger than 0.9 V/Å , with further increase of the field, the growth of the conduction population is becoming very small and looks like saturated. The growth rate of the conduction band population is consisted with the increase of the harmonic strength in the two lower regimes of the driving laser field. For the laser field stronger than 0.9 V/Å , the change of the growth rate was not observed in our experiment. Actually, the target was rapidly destroyed by the further increased laser field. For comparison the variation of the minimum band gap with the laser field is also shown in fig.4(a). The increase of the laser field not only brings the increase of the ionization from the valence band to conduction band, but also compress the bandgap of the solids which further pushes the increase of the ionization.

However, the minimum of the field modified band gap is encountered at 0.9 V/Å, which does not fit with the changing point of growth rate around 0.75 V/Å . To reveal the physics behind this critical changing point, the conduction band population is transformed from the moving frame $K$ to initial frame $k$(Fig.4c). For laser

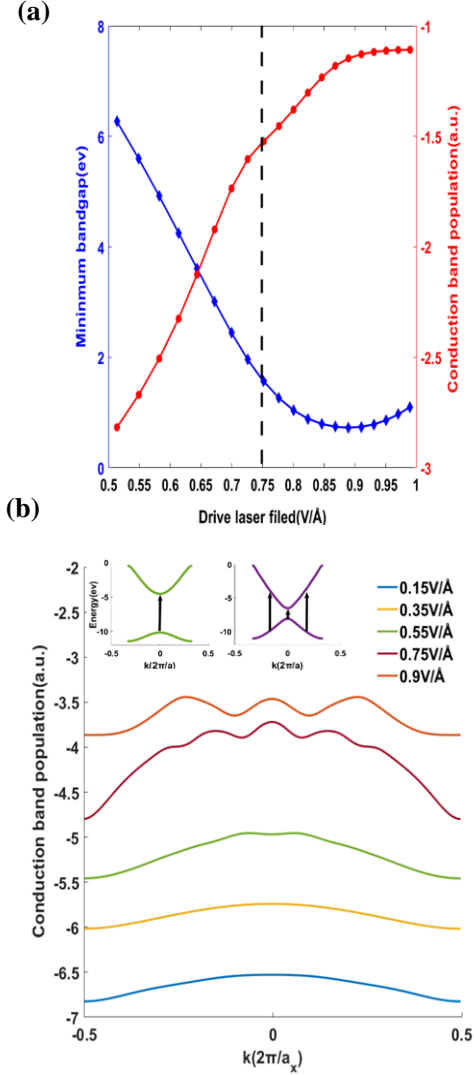

Figure 4 (a) The minimum energy bandgap (bule line) and total conduction band population (red line) simulated by band modified SBE versus drive laser field on a logarithmic scale for bulk $SiO_2$ (black dot). The dashed black line indicates that the minimum energy gap approaches the fundamental laser photon energy. (b) Conduction band population of different $k$ simulated by field-induced SBE at different laser field strengths in a logarithmic scale. When laser field strength is at $0.15\,\text{V/Å}$ (blue line), $0.35\,\text{V/Å}$ (yellow line), $0.55\,\text{V/Å}$ (green line), the conduction band population is mainly concentrated around $k = 0$ (the minimum bandgap). When laser field strength is at $0.75\,\text{V/Å}$ (purple line), $0.9\,\text{V/Å}$ (orange line), there are other peaks occur at $k \neq 0$. The inset figure shows the band structure at laser field $0.55\,\text{V/Å}$ (green line) and $0.75\,\text{V/Å}$ (purple line). The black arrows indicate the electron transitions.

field strength lower than $0.75\,\text{V/Å}$, the electron transitions mainly occur around the minimum bandgap. The probability of electron transition around minimum bandgap increases rapidly with quick rising field strength and at the same time the fast-falling bandgap, which leads to a rapidly rising peak of the conduction band population around the minimum bandgap (Fig.4b inset green line). For the laser field around $0.75\,\text{V/Å}$, the minimum bandgap decreases to $1.58\,\text{eV}$, which is the fundamental laser photon energy, in addition to the minimum bandgap, new transition channels start to open and electron population peaks emerge in other crystal momentum $k$, (Fig.4b inset purple line). For the laser field around $0.75\,\text{V/Å}$, new transition channels opens rapidly where the bandgap is about $3.7\,\text{eV}$. With the further increase of the laser field', the transition probability at minimum bandgap is suppressed. The growth rate of the population at the minimum bandgap slows down, and the other peaks at the non-minimum bandgap start to increase slowly, and the increase rate of the total electron population on the conduction band slows down (Fig.4a, green area). For the laser field intensity is close to $0.9\,\text{V/Å}$, the modified bandgap is the smallest. In this case, the height of other peaks of the non-minimum bandgap are same as the peak at the minimum bandgap, The total electron population in conduction band reaches the maximum.

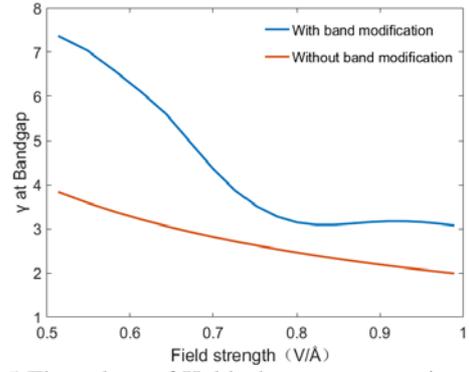

Fig.5 The values of Keldysh parameter $\gamma$ is calculated as a function of laser field strength at minimum bandgap with (bule line) and without (red line) band modification.

The nonperturbative ionization of solids by high intense laser field can be well described by Keldysh parameter, which is defined as $\gamma = \frac{w_0}{F_0}\sqrt{m_{e-h}E_g}$, where $E_g = E_c - E_v$ is the bandgap, $m_{e-h}$ is the effective-mass of electron-hole pair, $w_0$ is the laser frequency and $F_0$ is laser field strength. The calculated Keldysh parameter at the minimum bandgap is shown in Fig.5. Without considering the band modification, $\gamma$ continuously decreases in the whole field range, and no feature can be found to recognize the ionization regimes. The slowly decrease of $\gamma$ with the value close to 1 means the process is recognized as tunneling process in this model. With the band modification using our model, $\gamma$ rapidly decreases with the increase of the laser field intensity. A local minimum can be found around $0.8\,\text{V/Å}$ and remains around $3.1$ with the further increase of the laser field. The clear transition of $\gamma$ value indicates the transition of the ionization mechanism. Because the characteristics of the tunneling ionization is the formation of the potential barrier modified by intense laser field. Furthermore, the Keldysh parameter experiences a rapid decrease from 7 to 3 and keeps constant at 3. We can conclude that the mechanism of the ionization experienced the transition from multiphoton to tunneling in our experiment. According to the Keldysh theory, tunneling happened in our experiment should be diabatic tunneling, which is a process that the electron absorbs energy while moving under the barrier, combining the tunneling with multiphoton heating inside the classically forbidden region.

In conclusion, the power scale dependence of the high-order harmonic as a function of the driving laser field strength has been investigated. Two different slopes in the power scale dependence curve clearly divided the interaction into two regimes. Simple theoretic model shows it is important to include the field induced band modification in the SBE simulation. Only with the bandgap modification considered can the two interaction regimes be reproduced. Combining with the Keldysh theory analysis, we can conclude that, with the increase of the laser field intensity, the interaction process in our experiment was transitioned from multiphoton excitations to the diabatic tunneling. Adiabatic tunneling cannot be approached even with sub-10 femtosecond laser pulse. Our result bridges the gap between the non-perturbative HHG and the Keldysh description by considering the modified potential and bandgap, which is important for further application on the strong field solid physics investigation.

This work was supported by the National Natural Science Foundation of China (No.11974416).

Authors to whom correspondence should be addressed:
*xinkuihe@iphy.ac.cn;
#zywei@iphy.ac.cn

1. S. Ghimire, A. D. DiChiara, E. Sistrunk, P. Agostini, L. F. DiMauro and D. A. Reis, Nat. Phys. **7** (2), 138-141 (2011).
2. O. Schubert, M. Hohenleutner, F. Langer, B. Urbanek, C. Lange, U. Huttner, D. Golde, T. Meier, M. Kira, S. W. Koch and R. Huber, Nat. Photonics **8** (2), 119-123 (2014).
3. T. T. Luu, M. Garg, S. Y. Kruchinin, A. Moulet, M. T. Hassan and E. Goulielmakis, Nature **521** (7553), 498-502 (2015).
4. G. Vampa, T. J. Hammond, N. Thire, B. E. Schmidt, F. Legare, D. D. Klug, C. R. McDonald, T. Brabec, P. B. Corkum and Ieee, presented at the Conference on Lasers and Electro-Optics (CLEO), San Jose, CA, 2016 (unpublished).
5. G. Ndabashimiye, S. Ghimire, M. X. Wu, D. A. Browne, K. J. Schafer, M. B. Gaarde and D. A. Reis, Nature **534** (7608), 520-+ (2016).
6. M. Sivis, M. Taucer, G. Vampa, K. Johnston, A. Staudte, A. Y. Naumov, D. M. Villeneuve, C. Ropers and P. B. Corkum, Science **357** (6348), 303-306 (2017).
7. G. Vampa, T. J. Hammond, N. Thire, B. E. Schmidt, F. Legare, C. R. McDonald, T. Brabec, D. D. Klug and P. B. Corkum, Phys. Rev. Lett. **115** (19) (2015).
8. A. A. Lanin, E. A. Stepanov, A. B. Fedotov and A. M. Zheltikov, Optica **4** (5), 516-519 (2017).
9. L. Li, P. F. Lan, L. X. He, W. Cao, Q. B. Zhang and P. X. Lu, Phys. Rev. Lett. **124** (15), 6 (2020).
10. T. T. Luu and H. J. Worner, Nat. Commun. **9**, 6 (2018).
11. H. Z. Liu, Y. L. Li, Y. S. You, S. Ghimire, T. F. Heinz and D. A. Reis, Nat. Phys. **13** (3), 262-+ (2017).
12. Y. S. You, D. A. Reis and S. Ghimire, Nat. Phys. **13** (4), 345-349 (2017).
13. F. Langer, M. Hohenleutner, C. P. Schmid, C. Poellmann, P. Nagler, T. Korn, C. Schuller, M. S. Sherwin, U. Huttner, J. T. Steiner, S. W. Koch, M. Kira and R. Huber, Nature **533** (7602), 225-+ (2016).
14. D. Golde, T. Meier and S. W. Koch, Phys. Rev. B **77** (7), 6 (2008).
15. G. Vampa, C. R. McDonald, G. Orlando, P. B. Corkum and T. Brabec, Phys. Rev. B **91** (6) (2015).
16. M. X. Wu, S. Ghimire, D. A. Reis, K. J. Schafer and M. B. Gaarde, Phys. Rev. A **91** (4), 11 (2015).
17. E. N. Osika, A. Chacon, L. Ortmann, N. Suarez, J. A. Perez-Hernandez, B. Szafran, M. F. Ciappina, F. Sols, A. S. Landsman and M. Lewenstein, Physical Review X **7** (2) (2017).
18. G. Vampa, C. R. McDonald, G. Orlando, D. D. Klug, P. B. Corkum and T. Brabec, Phys. Rev. Lett. **113** (7) (2014).
19. T. Higuchi, M. I. Stockman and P. Hommelhoff, Phys. Rev. Lett. **113** (21) (2014).
20. L. J. S. P. J. Keldysh, **20** (5), 1307-1314 (1965).
21. H. Lakhotia, H. Y. Kim, M. Zhan, S. Hu, S. Meng and E. Goulielmakis, Nature **583** (7814), 55-+ (2020).
22. M. Garg, M. Zhan, T. T. Luu, H. Lakhotia, T. Klostermann, A. Guggenmos and E. Goulielmakis, Nature **538** (7625), 359-363 (2016).
23. T. T. Luu and H. J. Worner, Phys. Rev. A **98** (4), 5 (2018).
24. N. Tzoar and J. I. Gersten, Phys. Rev. B **12** (4), 1132-1139 (1975).
25. M. Gavrila and J. Z. Kaminski, Phys. Rev. Lett. **52** (8), 613-616 (1984).
26. M. Korbman, S. Y. Kruchinin and V. S. Yakovlev, New J. Phys. **15** (2013).